\documentclass[proof]{WileyASNA-v1}

\definecolor{darkblue}{cmyk}{1.0 , 0.5, 0.1, 0.1}
\definecolor{darkred}{cmyk}{0,1,1,0.45}
\hypersetup{colorlinks,breaklinks,
            linkcolor={red!60!black},urlcolor=darkblue,
            anchorcolor=Blue,citecolor=darkblue}

\usepackage{fontawesome}            
\definecolor{twitterblue}{RGB}{64,153,255}
\newcommand{\twitter}[1]{\protect\href{https://twitter.com/#1}{\textcolor{twitterblue}{\faTwitter}\,\textcolor{twitterblue}{@#1}}}            

\newcommand\blfootnote[1]{%
  \begingroup
  \renewcommand\thefootnote{}\footnote{#1}%
  \addtocounter{footnote}{-1}%
  \endgroup
}

\usepackage{twoopt,natbib}
\bibpunct{(}{)}{;}{a}{}{,} 
\makeatletter
\newcommandtwoopt{\citeads}[3][][]{\href{http://ui.adsabs.harvard.edu/abs/#3}%
{\def\hyper@linkstart##1##2{}%
\let\hyper@linkend\@empty\citealp[#1][#2]{#3}}}
\newcommandtwoopt{\citepads}[3][][]{\href{http://ui.adsabs.harvard.edu/abs/#3}%
{\def\hyper@linkstart##1##2{}%
\let\hyper@linkend\@empty\citep[#1][#2]{#3}}}
\newcommandtwoopt{\citetads}[3][][]{\href{http://ui.adsabs.harvard.edu/abs/#3}%
{\def\hyper@linkstart##1##2{}%
\let\hyper@linkend\@empty\citet[#1][#2]{#3}}}
\newcommandtwoopt{\citeyearads}[3][][]%
{\href{http://ui.adsabs.harvard.edu/abs/#3}
{\def\hyper@linkstart##1##2{}%
\let\hyper@linkend\@empty\citeyear[#1][#2]{#3}}}
\makeatother

\articletype{Conference Proceeding}%

\received{\today}
\revised{\today}
\accepted{\today}

\begin{document}

\title{Coronal Mass Ejections and Exoplanets: A Numerical Perspective\protect\thanks{Based on an invited review talk at the \href{https://www.cosmos.esa.int/web/xmm-newton/2021-workshop/}{XMM-Newton Science Workshop 2021: ``A High-Energy View of Exoplanets and Their Environments''}.}}

\author[1]{Juli\'an D. Alvarado-G\'omez*$^{,\ddagger,}$}

\author[2]{Jeremy J. Drake}

\author[3]{Ofer Cohen}

\author[2,4]{Federico Fraschetti}

\author[2,5]{\\ Cecilia Garraffo}

\author[1,6]{Katja Poppenh\"{a}ger}

\authormark{ALVARADO-G\'OMEZ \textsc{et al}}

\address[1]{\orgname{Leibniz Institute for Astrophysics (AIP)},\\ \orgaddress{\state{Potsdam}, \country{Germany}}}

\address[2]{\orgname{Smithsonian Astrophysical Observatory}, \orgaddress{\state{Cambridge, MA}, \country{USA}}}

\address[3]{\orgname{University of Massachusetts at Lowell}, \orgaddress{\state{Lowell, MA}, \country{USA}}}

\address[4]{\orgname{University of Arizona}, \orgaddress{\state{Tucson, AZ}, \country{USA}}}

\address[5]{\orgname{Harvard University}, \orgaddress{\state{Cambridge, MA}, \country{USA}}}

\address[6]{\orgname{University of Potsdam}, \orgaddress{\state{Potsdam}, \country{Germany}}}

\corres{*Juli\'an David Alvarado-G\'omez\\
\email{julian.alvarado-gomez@aip.de}\\
\twitter{AstroRaikoh}}

\presentaddress{Leibniz Institute for Astrophysics (AIP)\\An der Sternwarte 16, 14482 Potsdam, Germany}

\abstract{Coronal mass ejections (CMEs) are more energetic than any other class of solar phenomena. They arise from the rapid release of up to $10^{33}$ erg of magnetic energy mainly in the form of particle acceleration and bulk plasma motion. Their stellar counterparts, presumably involving much larger energies, are expected to play a fundamental role in shaping the environmental conditions around low-mass stars, in some cases perhaps with catastrophic consequences for planetary systems due to processes such as atmospheric erosion and depletion. Despite their importance, the direct observational evidence for stellar CMEs is almost non-existent. In this way, numerical simulations constitute extremely valuable tools to shed some light on eruptive behavior in the stellar regime. Here we review recent results obtained from realistic modeling of CMEs in active stars, highlighting their key role in the interpretation of currently available observational constraints. We include studies performed on M-dwarf stars, focusing on how emerging signatures in different wavelengths related to these events vary as a function of the magnetic properties of the star. Finally, the implications and relevance of these numerical results are discussed in the context of future characterization of host star-exoplanet systems.}





\maketitle

\blfootnote{\hspace{-0.275cm}$^\ddagger$Karl Schwarzschild Fellow}

\vspace{-1.3cm}
\section{State of the Art}\label{sec1}

The first spacecraft observations of coronal mass ejections (CMEs) on the Sun were performed in the early 1970s \citepads{1973BAAS....5..419T}. Since then, solar CMEs have been studied extensively allowing a detailed characterization of their physical properties, as well as their relation to solar flaring events (\citeads{2012LRSP....9....3W}, \citeads{2017LRSP...14....2B}). This relation is expected due to the fact that both forms of transient, flares and CMEs, originate from the sudden energy release involved in the reorganization of the coronal magnetic field (\citeads{2011LRSP....8....1C}, \citeads{2011LRSP....8....6S}). While solar flares and CMEs do not show a one-to-one correspondence, statistical analyses indicate that the CME association rate grows rapidly as the energy of the flare increases (e.g.~\citeads{2009IAUS..257..233Y}, \citeads{2017SoPh..292....5C}), and that the mass and kinetic energy of the emerging CME are positively correlated with the X-ray fluence of the flare (\citeads{2019SSRv..215...39L}).

In the context of exoplanets, flares and CMEs are of key importance for atmospheric characterization and any realistic assessment of habitability. The enhanced particle and photon fluxes associated with these events not only influence the chemistry of exoplanet atmospheres (e.g.~\citeads{2010AsBio..10..751S}, \citeads{2019AsBio..19...64T}), but also could drive catastrophic exoplanet mass loss and atmospheric depletion processes (\citeads{2013oepa.book.....L}, \citeads{2017ApJ...846...31C}). These effects are expected to be exacerbated for planets in the close-in habitable zones (HZ) of M-dwarfs, as these stars remain magnetically-active for a much longer period of time compared to any other spectral type (\citeads{2019ApJ...871..241D}, \citeads{2021A&A...649A..96J}).

The flaring behaviour of cool stars is nowadays well documented, with a large number of events detected in simultaneous multi-wavelength observations (e.g.~\citeads{2019A&A...622A.210G}, \citeads{2021ApJ...911L..25M}), extending across spectral types (e.g.~\citeads{2014ApJ...792...67C}, \citeads{2021MNRAS.504.3246J}) and evolutionary stages (e.g.~Ilin et al.~\citeyearads{2019A&A...622A.133I}, \citeyearads{2021A&A...645A..42I}, \citeads{2021ApJ...906...40G}). This increasing amount of information provides the foundation to study the effects of flares on exoplanet atmospheres and habitability (e.g.~\citeads{2018ApJ...865..101M}, \citeads{2021NatAs...5..298C}). In stark contrast, knowledge of the influence of stellar CMEs on exoplanetary systems is still in very early stages, with most literature relying on direct solar extrapolations (e.g.~\citeads{2017SoPh..292...89P}, \citeads{2019ApJ...886L..37K}). It is worth noting that in the case of the Sun, flares and CMEs fare equally --perhaps even the latter dominate-- the energetic budget of magnetically-driven transients (\citeads{2012ApJ...759...71E}, \citeads{2017ApJ...836...17A}). Therefore, a proper assessment of the consequences of these space weather events on exoplanets must include the contribution from stellar CMEs.  

It is only natural to assume that flares and CMEs on stars with similar magnetic activity level and spectral type as the Sun should behave similarly to solar events. On the other hand, \citetads{2013ApJ...764..170D} showed that the solar relationships, connecting the flare X-ray fluence with the CME mass and kinetic energy, cannot be extended to the most magnetically active stars --typically characterized by enhanced flare rates and energies (see e.g.~\citeads{2002ApJ...580.1118K}, \citeads{2013ApJS..209....5S}, \citeads{2018ApJ...867...71L}). This is because the mass loss and energy requirements resulting from the expected CME activity become unreasonably large (see also \citeads{2015ApJ...809...79O}, \citeads{2017MNRAS.472..876O}). 

The inference from \citetads{2013ApJ...764..170D} is also backed up by the current observational picture of stellar CMEs. While at face value the increased flare rates and energies of active stars should correspond to high levels of eruptive activity, this is in fact at odds with the consistent lack of CME detections in dedicated observations (e.g.~\citeads{2016ApJ...830...24C}, \citeads{2018ApJ...856...39C}, \citeads{2019A&A...623A..49V}, Muheki et al.~\citeyearads{2020A&A...637A..13M}, \citeyearads{2020MNRAS.499.5047M}). So far, the sole direct detection of a stellar CME was obtained by \citetads{2019NatAs...3..742A}, identified from blue-shifted cool coronal emission following a flare on the giant star HR\,9024 (G1III). The estimated properties for this event ($M_{\rm CME} \sim 10^{21}$~g, $E^{\rm K}_{\rm CME} \sim 10^{34}$~erg) exceed the nominal parameter space observed in solar eruptions ($M_{\rm CME} \lesssim 10^{17}$~g, $E^{\rm K}_{\rm CME} \lesssim 10^{33}$~erg). Note that, while this CME develops under a magnetic and coronal environment comparable with the expectation for active Sun-like stars (i.e., $B_{\bigstar} \sim 200-400$~G, $L_{\rm X} \sim 10^{31}$~erg~s$^{-1}$, \citeads{2016A&A...591A..57B}), the surface gravity of its host star is significantly reduced (by a factor of $\sim30$).

Recently, \citetads{2021NatAs.tmp...72V} reported a series of 21 coronal dimmings on 13 active late-type stars which the authors interpreted as signatures of CMEs. Unfortunately, this study did not provide estimates for any of the physical properties of these events. Previously, \citetads{2019ApJ...877..105M} revisited the best stellar flare-CME candidates in the literature, obtaining constraints on CME parameters such as mass and kinetic energy. Their analysis included events identified from continuous X-ray absorption following a stellar flare (e.g.~\citeads{1999A&A...350..900F}, \citeads{2017ApJ...850..191M}), as well as candidates from post-flare Doppler-shift signatures in optical or UV spectral lines (e.g.~\citeads{1990A&A...238..249H}, \citeads{2011A&A...536A..62L}). 
The compilation of \citetads{2019ApJ...877..105M} revealed fundamental hints of the behaviour of the flare-CME relation in the stellar domain. While the estimated CME masses appeared roughly aligned with a solar extrapolation to the observed X-ray energy of the flares, the kinetic energy of each eruption was significantly lower than the expectation. The properties of the CME inferred by \citetads{2019NatAs...3..742A} follow the same trend.

Our recent numerical studies have proven to be critical to interpret these observations and understand the apparent imbalance between flare and CME occurrence in stars. Using solar-validated 3D Magneto-Hydrodynamic (MHD) models, we showed that the stellar large-scale magnetic field can partially or totally suppress CMEs of certain energies from escaping (see \citeads{2016IAUS..320..196D}, \citeads{2018ApJ...862...93A}). Apart from providing a pathway out of the quandary identified by \citetads{2013ApJ...764..170D}, our results successfully predicted the properties of the best stellar CMEs observed so far. Furthermore, our investigation has also revealed various consequences of this CME magnetic-suppression process in active M-dwarf stars including novel observational signatures in different wavelengths (Alvarado-G\'omez~et~al.~\citeyearads{2019ApJ...884L..13A}, \citeyearads{2020ApJ...895...47A}). 

Here we present an overview of our numerical investigation of stellar CMEs. A description of the codes and methodology is provided in Sect.~\ref{sec2}. Section \ref{sec3} contains a brief summary and highlights from some of our previous studies. Conclusions and perspectives in this area are given in Sect.~\ref{sec4}.

\begin{figure*}[!t]
\centering
\includegraphics[trim=0.0cm 0.0cm 0.0cm 0.0cm, clip=true, width=0.33\textwidth]{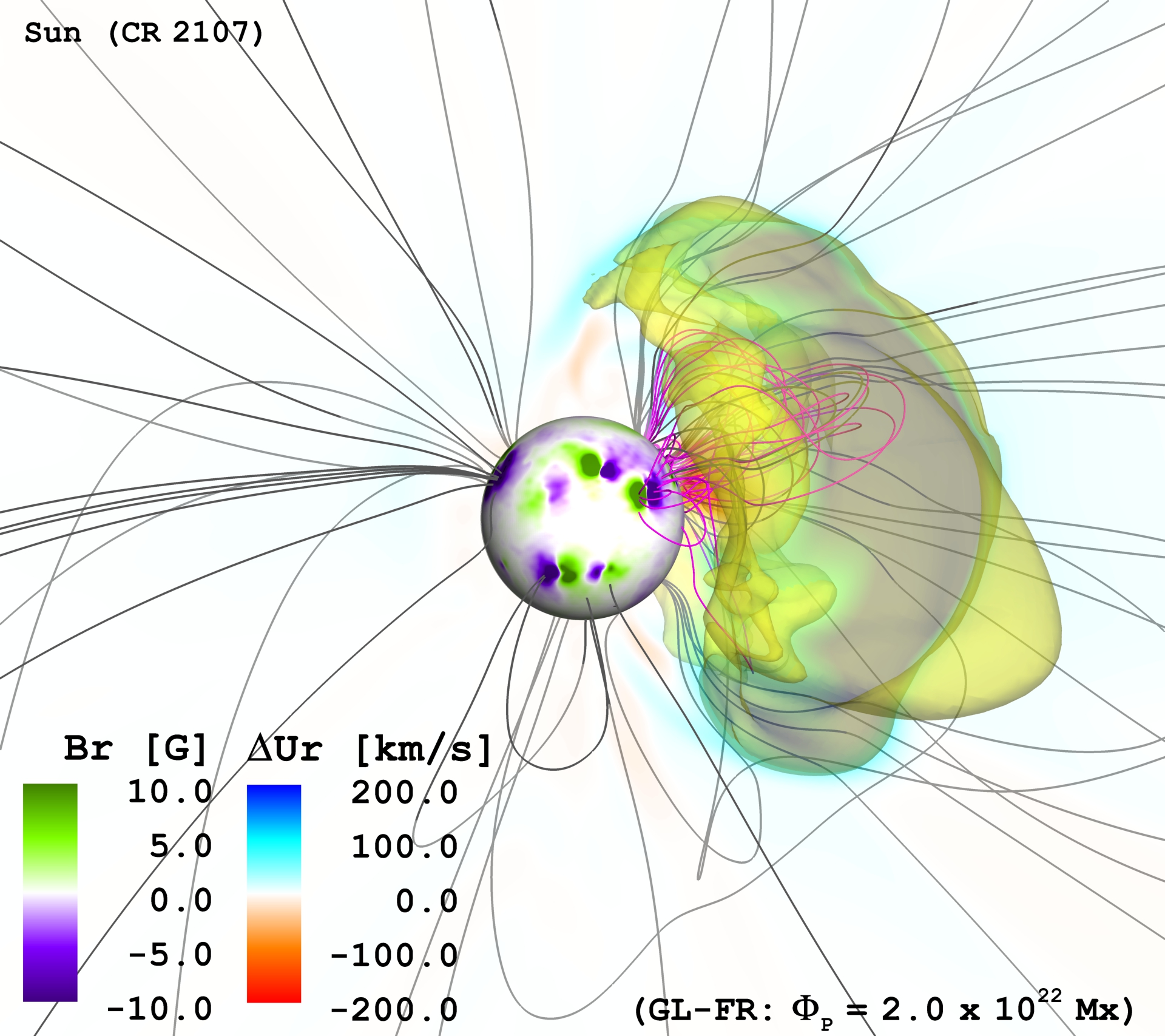}\hspace{1pt}\includegraphics[trim=0.0cm 0.0cm 0.0cm 0.0cm, clip=true, width=0.33\textwidth]{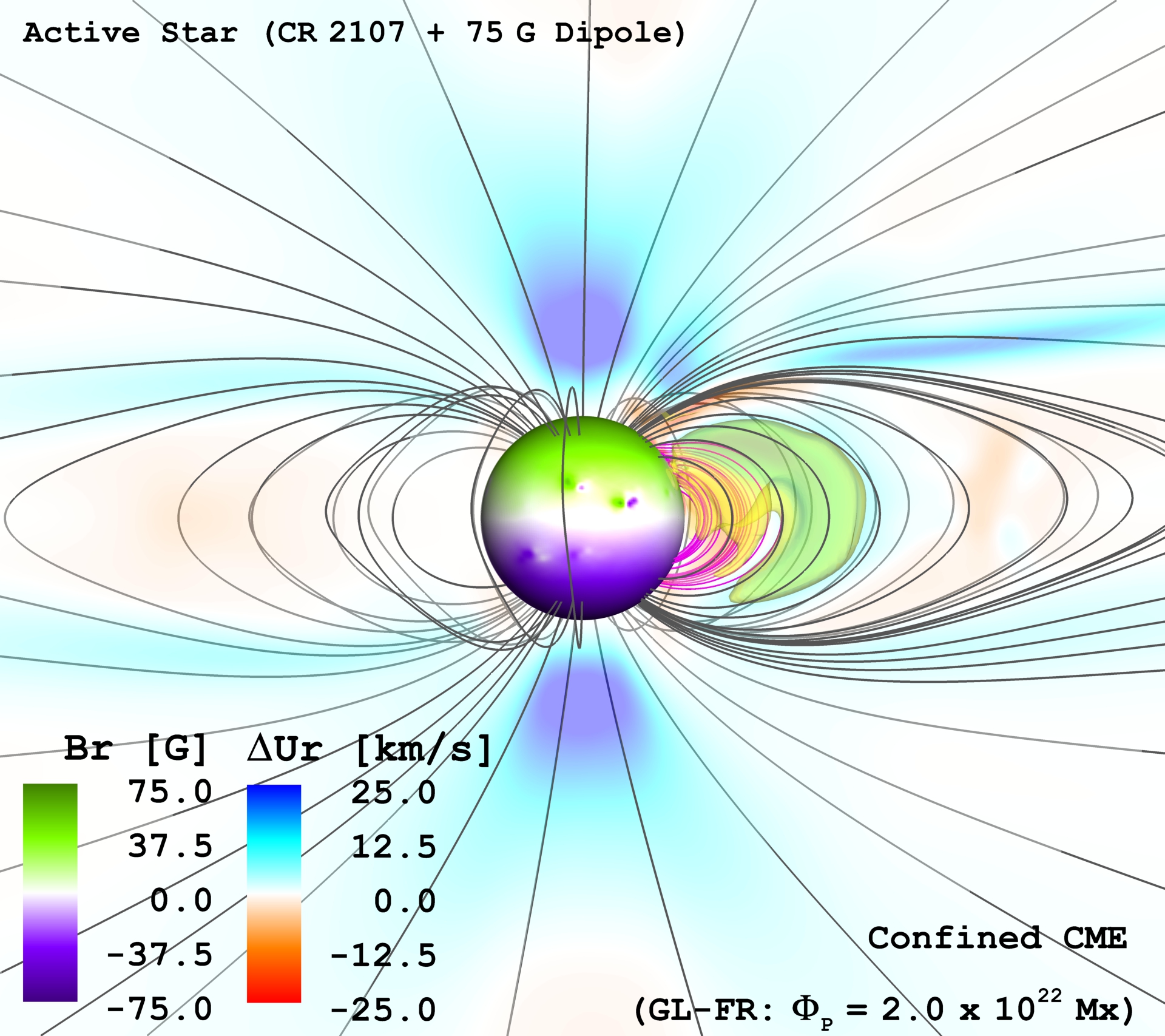}\hspace{1pt}\includegraphics[trim=0.0cm 0.0cm 0.0cm 0.0cm, clip=true, width=0.33\textwidth]{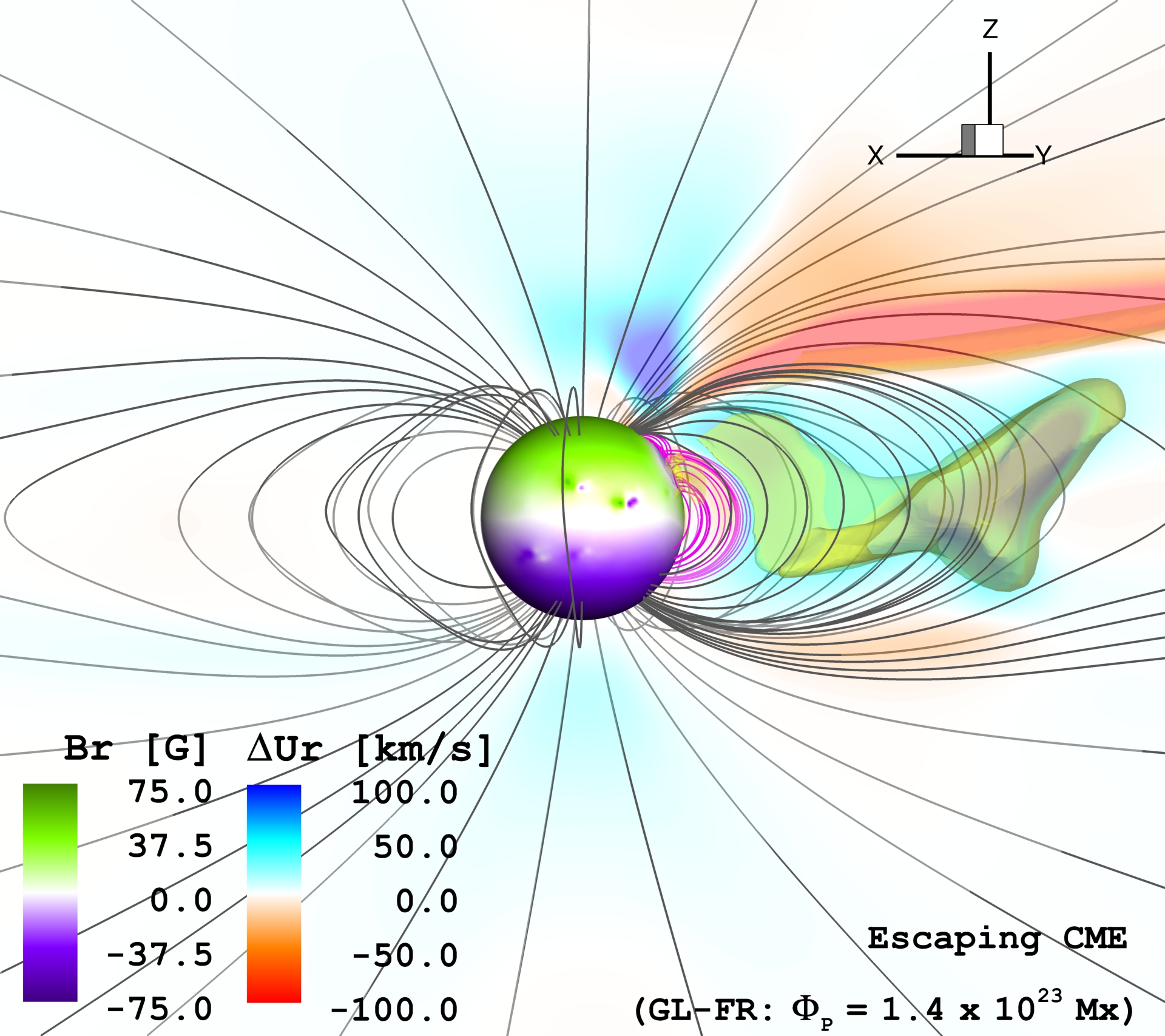}
\caption{Results from different AWSoM+GL flux-rope simulations employed to investigate the suppression of CMEs in active stars (see text for details). The stellar surface (central sphere) is colored by the magnetic field driving each model. The coronal Doppler-shift radial velocities ($\Delta U_{\rm r}$), calculated with respect to the pre-CME conditions, are indicated by a secondary color scale. A translucent yellow iso-surface identifies the eruption in each case. Streamlines in magenta and grey are associated to the eruptive active region and the large-scale magnetic field, respectively. The field of view in all panels is 12~$R_{\bigstar}$.}\vspace{-10pt}
\label{Fig_1}
\end{figure*}

\vspace{-0.5cm}
\section{Numerical Methodology}\label{sec2}

The results discussed in the following sections have been obtained using the open-source Space Weather Modeling Framework (SWMF, \citeads{2018LRSP...15....4G}). The SWMF encompasses a set of numerical models and tools developed to study the space environment of the Sun, which have been since adapted to study astrophysical systems (e.g.~\citeads{2014ApJ...790...57C}, \citeads{2020ApJ...902L...9A}). 

Three different modules of the SWMF have been considered in our investigation of stellar CMEs. The first one is the Alfv\'en Wave Solar Model (AWSoM, \citeads{2014ApJ...782...81V}), which uses the radial magnetic field on the solar/stellar surface as a boundary condition to self-consistently calculate the coronal heating and stellar wind acceleration due to Alfv\'en wave turbulent dissipation. Solar chromospheric values of plasma density ($n_{0} = 2\times10^{10}$~cm$^{-3}$) and temperature ($T_{0} = 5\times10^{4}$~K) complement the starting parameters of the simulation. This module provides a steady-state description of the quiescent stellar wind and coronal structure (see also \citeads{2021ApJ...908..172S}). 

Once a quiescent solution is obtained, time-dependent flux-rope eruption models are coupled to the AWSoM domain to simulate a CME event. These correspond to numerical implementations of the \citetads[GL,]{1998ApJ...493..460G} and \citetads[TD,]{1999A&A...351..707T} configurations. 
While each model has its own set of initialization parameters, in both cases an unstable twisted flux-tube is anchored to a mixed-polarity region in the inner boundary of the simulation. While the energy build-up is not considered within the AWSoM+GL/TD models, the CME dynamics and evolution within the corona and stellar wind are properly captured in the simulations (see Jin et al.~\citeyearads{2013ApJ...773...50J}, \citeyearads{2017ApJ...834..173J}). 

\vspace{-0.5cm}
\section{A numerical view of Stellar Coronal Mass Ejections}\label{sec3}

\subsection{Suppression of CMEs in active stars}\label{sec3.1}

Spectropolarimetry and the technique of Zeeman-Doppler Imaging (ZDI) have revealed an ubiquitous presence of magnetic fields in cool main-sequence stars (\citeads{2009ARA&A..47..333D}, \citeads{2012LRSP....9....1R}, \citeads{2021A&ARv..29....1K}). Sensitive to the large-scale components, ZDI observations of active stars indicate surface field strengths much larger than their equivalent on the Sun (e.g.~\citeads{2012A&A...540A.138M}, \citeads{2017MNRAS.471L..96J}, \citeads{2019ApJ...876..118S}). These differences correspond to photospheric large-scale magnetic fields in the range of $10$~G to several $10^{2}$~G in young Sun-like stars (e.g.~\citeads{2015A&A...582A..38A}, Folson et al.~\citeyearads{2016MNRAS.457..580F}, \citeyearads{2018MNRAS.474.4956F}), and up to a few $10^{3}$~G for M-dwarfs (e.g.~\citeads{2010MNRAS.407.2269M},  \citeads{2018MNRAS.479.4836L}).  

\citetads{2016IAUS..320..196D} suggested that a sufficiently strong large-scale magnetic field could confine CMEs unless their energy would surpass an escaping threshold. This idea is motivated by solar observations, where confined eruptions are studied in the context of flare-rich CME-poor active regions (e.g.~\citeads{2015ApJ...804L..28S}, \citeads{2016ApJ...826..119L}). To test this hypothesis, we ran a series of numerical simulations employing the AWSoM+GL models described in Sect.~\ref{sec2}. Figure~\ref{Fig_1} shows the results of three separate cases illustrating the magnetic confinement of CMEs in active stars.

\begin{figure*}[!t]
\centering
\includegraphics[trim=0.0cm 0.0cm 0.0cm 0.0cm, clip=true, height=0.315\textwidth]{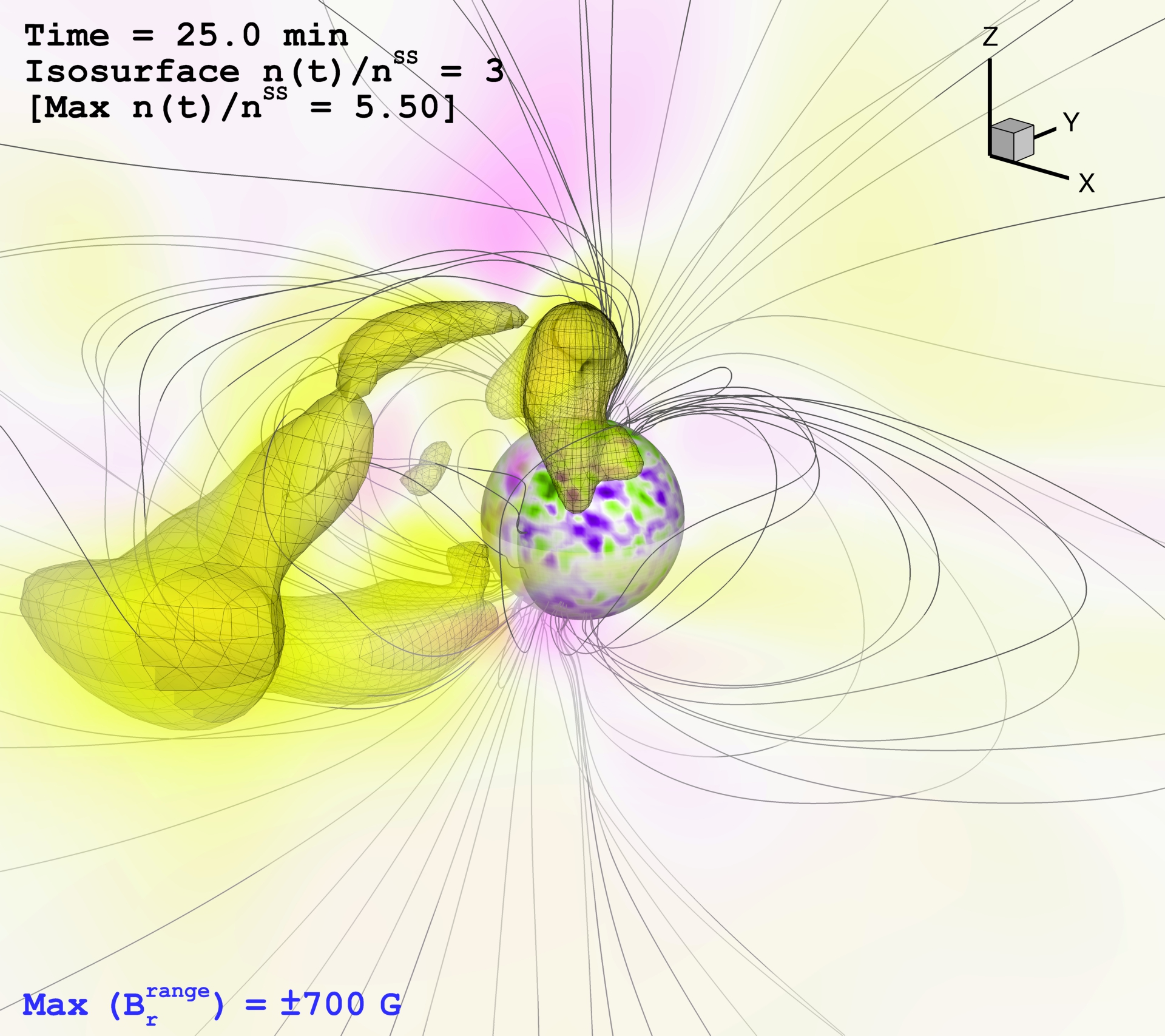}\hspace{1pt}\includegraphics[trim=0.0cm 0.0cm 0.0cm 0.0cm, clip=true, height=0.315\textwidth]{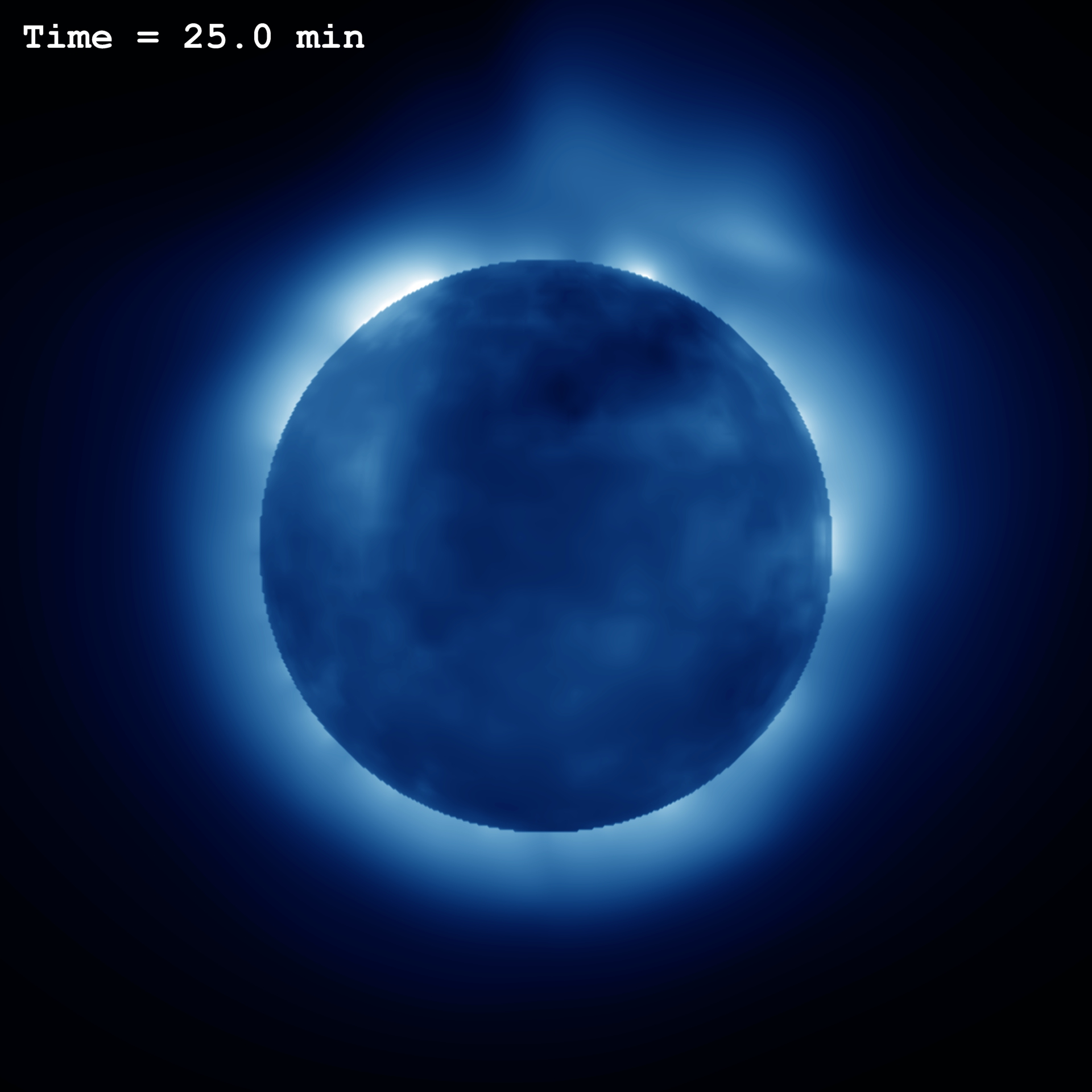}\hspace{1pt}\includegraphics[trim=0.0cm 0.0cm 0.0cm 0.0cm, clip=true, height=0.315\textwidth]{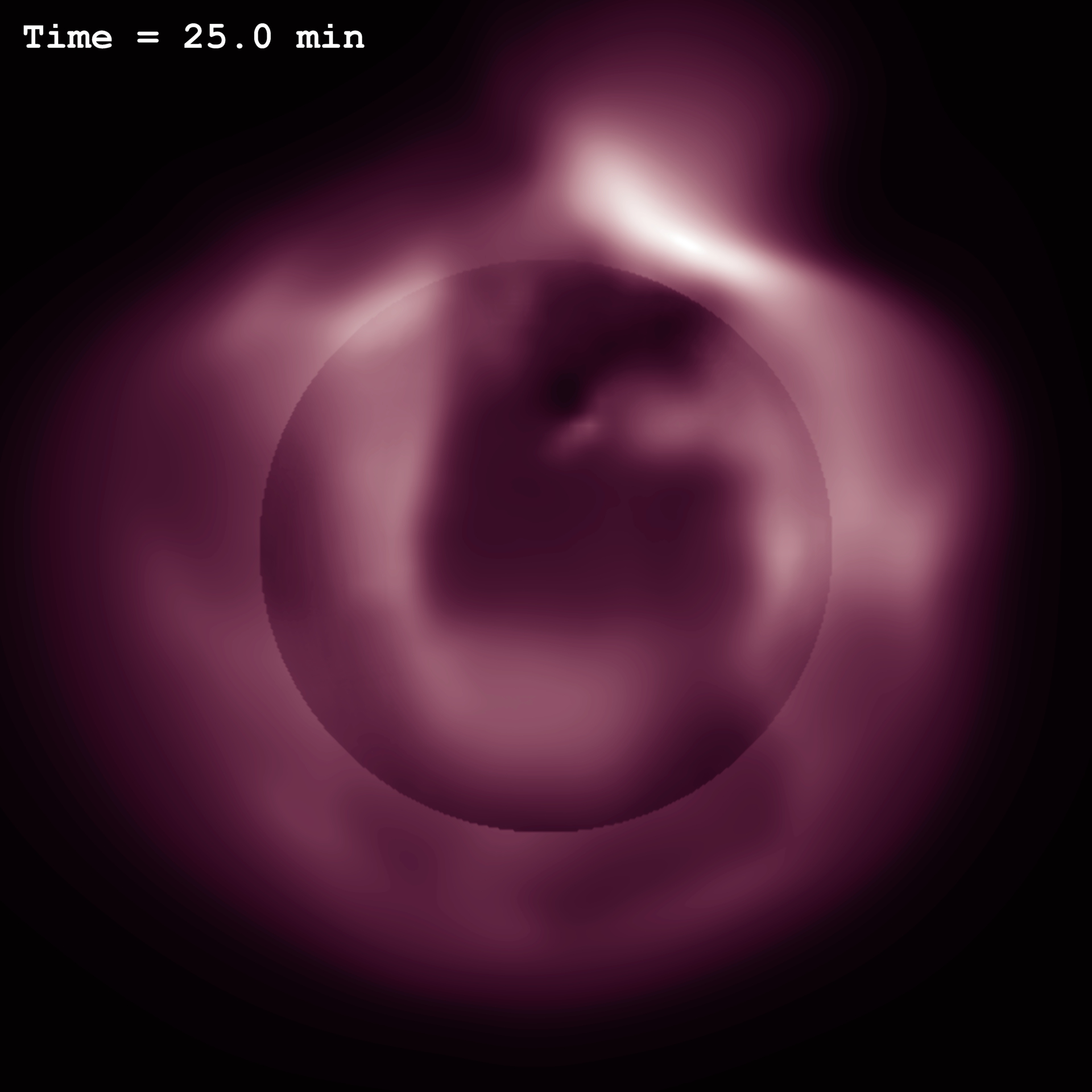}\vspace{1pt}
\includegraphics[trim=0.0cm 0.0cm 0.0cm 0.0cm, clip=true, height=0.315\textwidth]{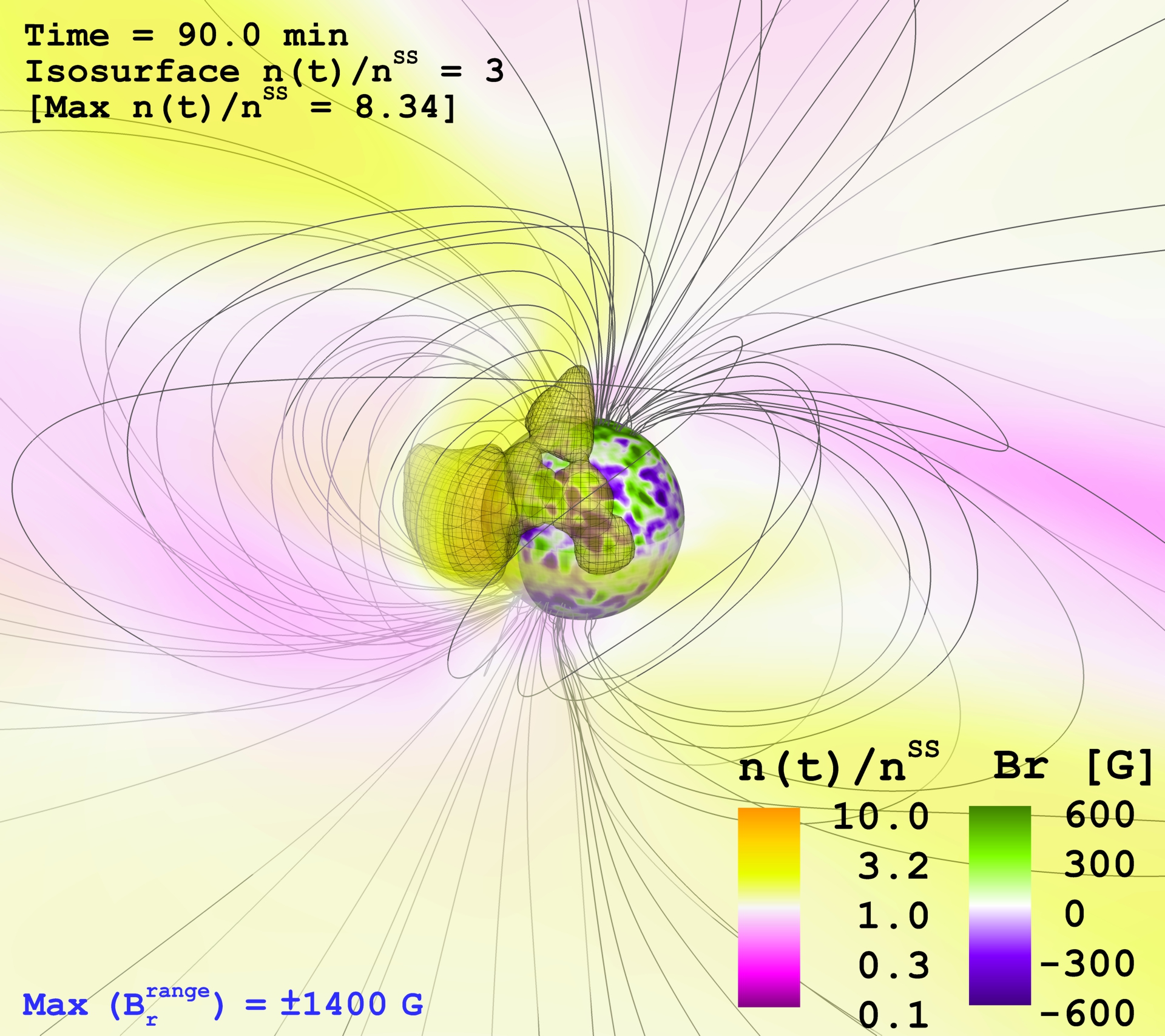}\hspace{1pt}\includegraphics[trim=0.0cm 0.0cm 0.0cm 0.0cm, clip=true, height=0.315\textwidth]{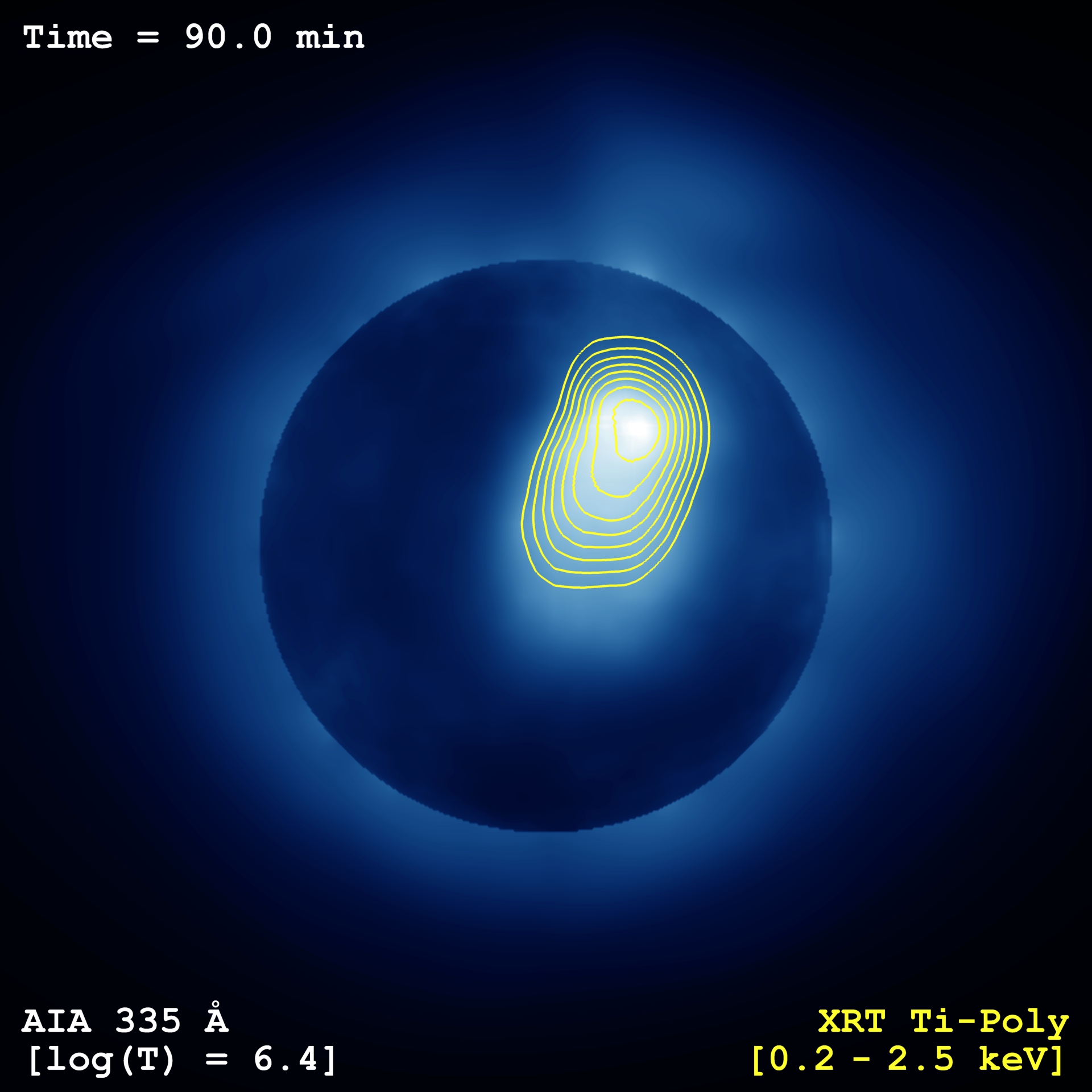}\hspace{1pt}\includegraphics[trim=0.0cm 0.0cm 0.0cm 0.0cm, clip=true, height=0.315\textwidth]{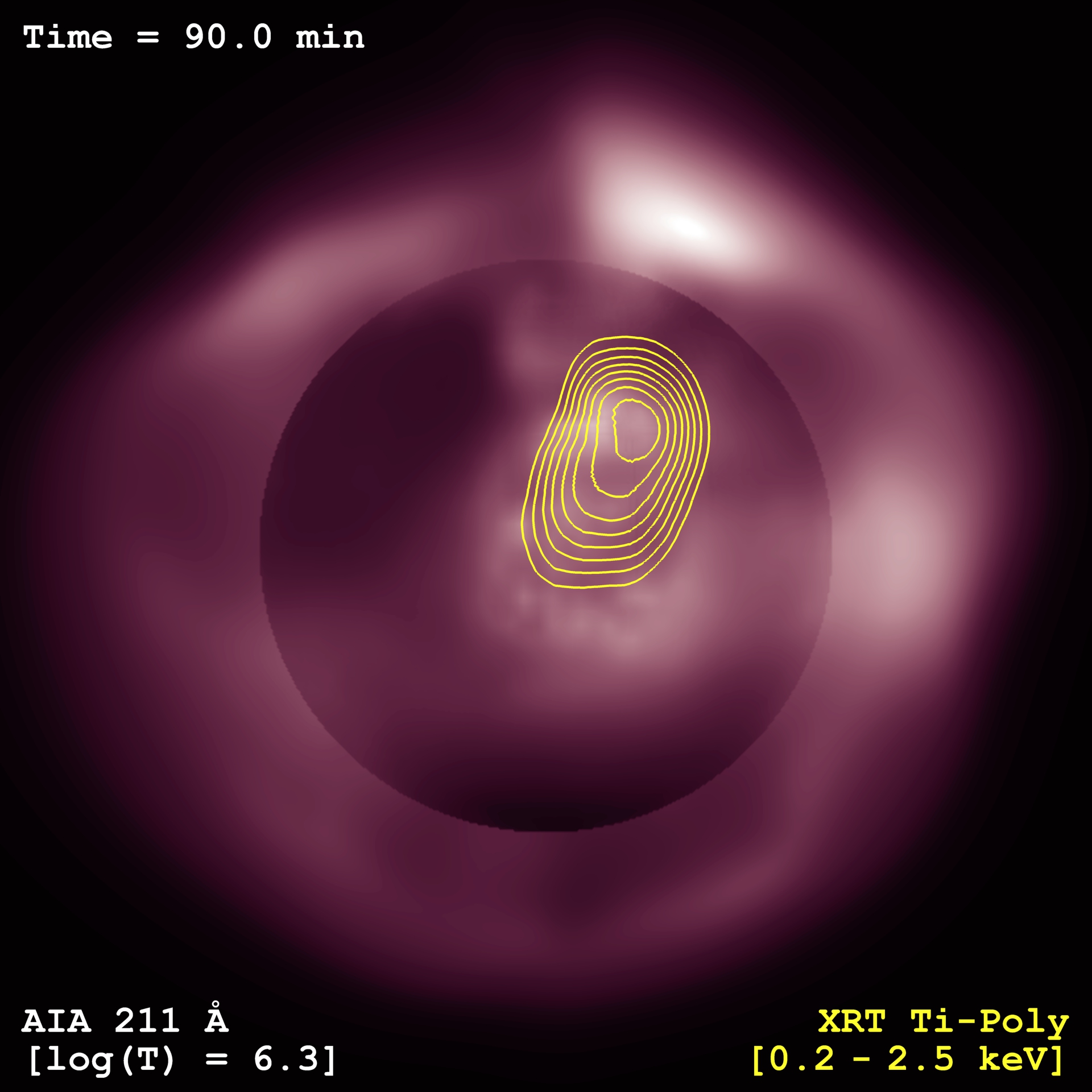}
\caption{Snapshots from two time-dependent numerical simulations of CMEs on M-dwarf stars. The geometry of the magnetic field driving AWSoM is provided by the 3D dynamo simulation performed by \citetads{2016ApJ...833L..28Y}, but each case uses a different scaling for the maximum surface field (\textit{Top:}~$B_{\rm r}^{\rm max}\,= \pm\,700$\,G, \textit{Bottom:}~$B_{\rm r}^{\rm max}\,=\,\pm\,1400$\,G). While the same TD flux-rope eruption is launched in both cases, the CME is only able to escape in the weaker magnetic configuration (as evidenced by the behaviour of the density contrast $n(t)/n^{\rm SS}$ visualised on the \textit{left} panels). The \textit{middle} and \textit{right} panels show the response of the coronal structure to each event. The latter is captured by simulated full-disk images (observer located towards the negative $y$ direction) synthesised in different bands of the XUV emulating solar instrumentation (AIA/SDO and XRT/Hinode). 
}\vspace{-10pt}
\label{Fig_2}
\end{figure*}

The left panel of Fig.~\ref{Fig_1} shows the simulation of a large solar CME ($M_{\rm CME} \sim 10^{17}$~g, $E^{\rm K}_{\rm CME}~\sim~10^{32}$~erg), occurring in one of the active regions of the Carrington Rotation (CR) 2107. This event is triggered by an erupting GL poloidal magnetic flux of $\Phi_{\rm p} \simeq 2.0 \times 10^{22}$~Mx --an equivalent amount of reconnecting flux would be associated with an X5.0 solar flare in the GOES classification (see \citeads{2018ApJ...862...93A}). The same CME appears confined when the eruption occurs under the influence of a $75$~G large-scale dipole field ($\sim$\,$35$ times larger than the solar one\footnote[1]{Assuming an (overestimated) average $2$~G large-scale dipole field for the Sun (see~\citeads{2015ApJ...798..114S}).}) aligned with the stellar rotation axis (Fig.~\ref{Fig_1}, middle). Even though this magnetic configuration is idealized, it is certainly plausible for $\sim$\,$0.4 - 0.8$ Gyr old F-G-K main sequence stars (e.g.~\citeads{2014A&A...569A..79J}, \citeads{2016A&A...585A..77H}). The confinement imposed by this configuration is relatively strong, arresting flux-rope eruptions up to the equivalent reconnecting flux of an $\sim$\,X20 solar flare. Still, more energetic events will be able to overcome the large-scale magnetic suppression. This is presented in the right panel of Fig.~\ref{Fig_1}, where an escaping CME is generated ($M_{\rm CME} \simeq 4.1 \times 10^{17}$~g, $E^{\rm K}_{\rm CME}\simeq 6.4 \times 10^{33}$~erg) after the eruption of a GL flux-rope with a larger poloidal flux of $\Phi_{\rm p} \simeq 1.4 \times 10^{23}$~Mx (equivalent to a $\sim$\,X150 GOES solar flare). If this event would have been simulated in a nominal solar magnetic field configuration, its expected kinetic energy would have been almost two orders of magnitude larger (i.e., $E^{\rm K}_{\rm CME}\simeq 2.3 \times 10^{35}$~erg, following the scaling obtained by \citeads{2017ApJ...834..173J}).

In general, our simulation results revealed that the overlying field would significantly slow down the escaping CMEs (reducing their associated kinetic energies) in contrast with extrapolations from solar data. Similarly, we found that the total mass perturbed in confined and escaping events roughly followed the solar flare-CME relation extended to the stellar regime. As discussed in Sect.~\ref{sec1}, both of these predictions have been corroborated by the current observational constraints of stellar CMEs (\citeads{2019ApJ...877..105M}, \citeads{2019NatAs...3..742A}).

\vspace{-0.4cm}
\subsection{Observational signatures of suppressed CMEs: The case of M-dwarf stars}\label{sec3.2}

Being the most abundant and long-lived stars in the galaxy, M-dwarfs are currently among the primary targets in exoplanet searches. This is largely due to their small size which increases the likelihood of detecting orbiting Earth-sized planets in the HZ with transit techniques, or due to their low mass compared with other spectral types which increases planet-induced radial velocity Doppler shifts in the stellar spectra \citepads{2018haex.bookE....D}. Their reduced bolometric luminosity implies that the temperature-based HZ, where liquid water can be sustained, is located very close to the central star --as much as ten or more times closer than the Sun-Earth distance (e.g.~\citeads{2014ApJ...787L..29K}, \citeads{2016PhR...663....1S}). These close-in orbits have short periods --from a few days up to a week-- facilitating the detection and confirmation of planet candidates.

On the other hand, M-dwarf stars are known for frequent and energetic flaring activity (e.g.~\citeads{2018ApJ...867...71L}, \citeads{2020AJ....159...60G}) even at relatively old ages in their evolution \citepads{2020AJ....160..237F}. Additionally, as mentioned in the last section, the strongest magnetic fields reported amongst main-sequence stars --up to a few kG in magnitude-- have been detected precisely in this spectral type. This has been confirmed by other magnetic field diagnostics such as Zeeman broadening (\citeads{2019A&A...626A..86S}) and Zeeman intensification of spectral lines (\citeads{2020ApJ...902...43K}). 

All these reasons make the numerical investigation of CMEs in M-dwarfs, including their possible confinement, of high relevance for stellar and exoplanet science. In particular, simulations can provide guidelines and constraints for current and future observational searches of these energetic transients. 

\smallskip

\noindent\textbf{CME-induced coronal signatures:}

\noindent Following this scientific rationale, Fig.~\ref{Fig_2} contains results from the first realistic modelling of CME events on M-dwarf stars \citepads{2019ApJ...884L..13A}. These simulations follow the methodology explained in Sect.~\ref{sec2}, with adjustments in the stellar parameters (e.g.\ mass, radius, rotation period) suitable for this spectral type. Furthermore, the geometry of the stellar magnetic field driving the AWSoM solutions is now provided by predictions from a 3D dynamo model of the M-dwarf star Proxima Centauri \citep{2016ApJ...833L..28Y}. We scaled the field strength at the photosphere to reach average values comparable to low- and moderately-active M-dwarf stars (\citeads{2014IAUS..302..156R}). The cases included in Fig.~\ref{Fig_2} correspond to magnetic field scalings between $\pm700$~G (top) and $\pm1400$~G (bottom). 

As shown in the 3D visualizations, while an identical TD flux-rope eruption is initiated in both cases ($M_{\rm FR} = 4\,\times\,10^{17}$~g, $E_{\rm FR}^{\rm B,\,free} = 10^{35}$ erg) an escaping CME develops only for the case with a relatively weak surface magnetic field (top-left panel). In the stronger magnetic field case the CME is totally confined by the large-scale magnetic field (bottom-left panel). Naturally, the coronal response is completely different in both cases as evidenced by the simulated full-disk images (in arbitrary normalized units) of the middle and right panels. These have been synthesized emulating the response of various solar instruments observing in EUV and X-ray wavelengths. From the perspective of an exoplanet orbiting the star, both events could strongly influence its atmosphere; either from the high dynamic pressure carried by the passing CME (top) or by the excess of high-energy emission induced by the confined event (bottom). A quantitative analysis of the various coronal signatures due to weakly, partially, and strongly confined CMEs in M-dwarfs, as well as their detectability prospects with current instrumentation can be found in \citetads{2019ApJ...884L..13A}. 

\smallskip
\noindent\textbf{Type II radio bursts and CME suppression:}

\begin{figure}[!b]
\vspace{-0.5cm}
\centering
\includegraphics[trim=0.0cm 0.0cm 0.0cm 0.0cm, clip=true, width=0.49\textwidth]{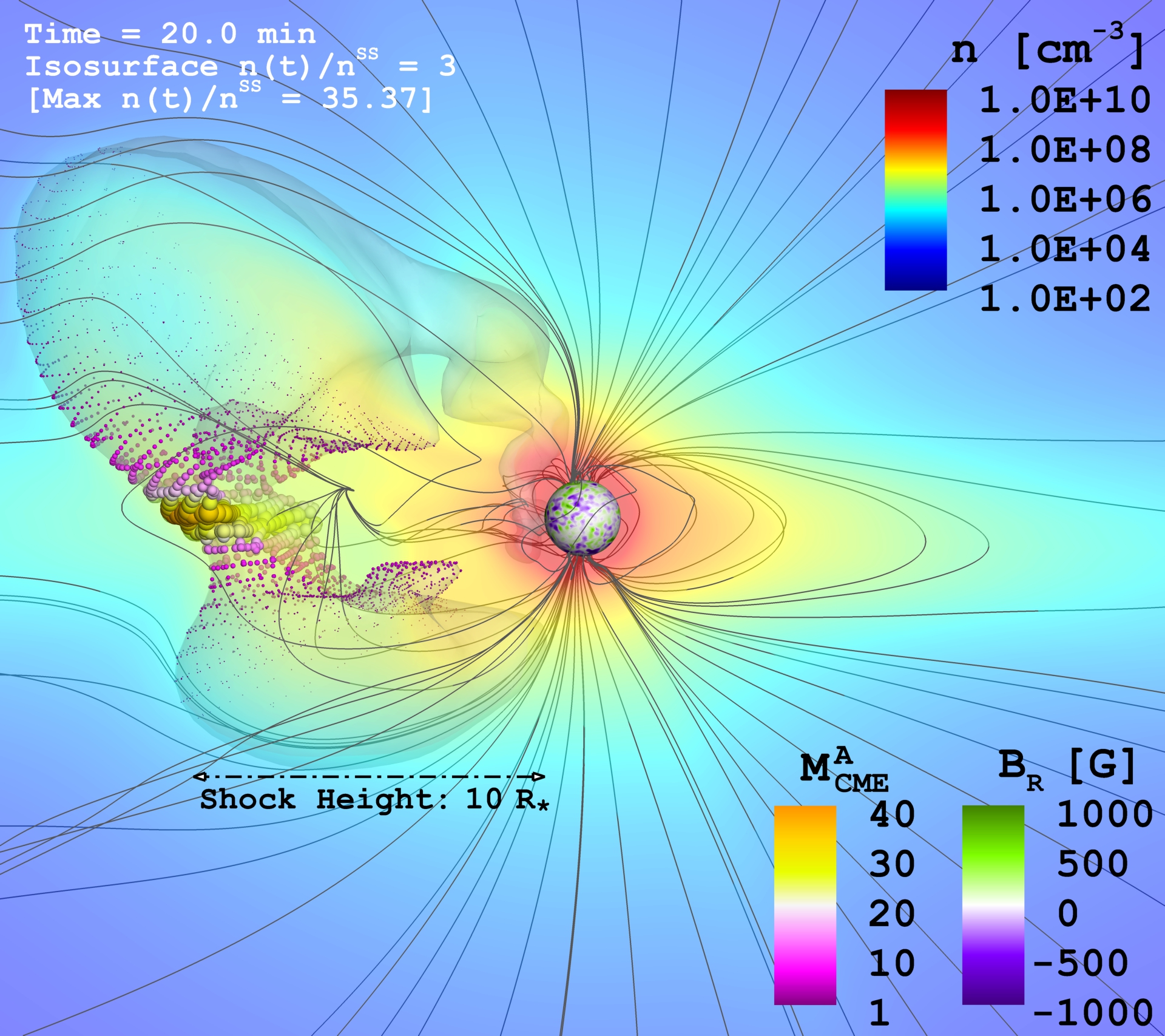}
\caption{Time-dependent numerical simulation of a CME in the flare star Proxima Centauri. The simulation setup is similar to the one presented in Fig.~\ref{Fig_2}. The identified front of the escaping CME (isosurface $n(t)/n^{\rm SS} =~3.0$) is overlaid on the distribution of the particle density in the corona ($n$~[cm$^{-3}$]). The snapshot corresponds to the moment in which a shock is generated at the CME front. This is quantified by the Alfv\'enic Mach number of the CME ($M_{\rm CME}^{\rm A}$), which is encoded by the size of the scatter distribution (spheres) and by an additional color-scale. The initial shock height generated by this weakly confined CME is approximately $10~R_{\bigstar}$ as indicated.       
}\vspace{-10pt}
\label{Fig_3}
\end{figure}

\noindent In a follow up study, we considered the consequences due to magnetic suppression on the expected Type II radio bursts emerging from escaping CMEs in M-dwarf stars \citepads{2020ApJ...895...47A}. Radio bursts of Type~II are indicative of an MHD shock occurring in the corona and interplanetary space, that give rise to electron acceleration and subsequent radio emission at the plasma frequency\footnote[2]{Defined (in c.g.s. units) as $\nu_{\rm p} = (2\pi)^{-1}(\sqrt{4\pi e^2/m_{\rm e}})\sqrt{n}$, where $m$, $e$ are the electron mass and charge, and $n$ denotes the particle density of the region.} and its first harmonic (see~\citeads{2003SSRv..107...27C} and references therein). These shocks are generated when the velocity of a CME becomes larger than the local Alfv\'en speed\footnote[3]{Given by $V^{\rm A} = B/\sqrt{4\pi \rho}$ with $B$ and $\rho$ as the local values of the magnetic field strength and the plasma density, respectively.}, a transition that on the Sun might occur at relatively low heights above the surface ($H\sim2-3~R_{\odot}$, \citeads{2005JGRA..11012S07G}, Ramesh et al.~\citeyearads{2012ApJ...752..107R}). Type~II radio bursts are therefore very good tracers of energetic solar CMEs (e.g.~Gopalswamy et al.~\citeyearads{2009SoPh..259..227G}, \citeyearads{2019SunGe..14..111G}). As such, and while unsuccessfully, they have been long sought-after in the context stellar CMEs (\citeads{2017PhDT.........8V}, \citeads{2017IAUS..328..243O}).

As the appearance of Type II radio bursts depends on CME speeds, it is clear that these radio transients will be directly affected by the suppression of CMEs in active stars (see Sect.~\ref{sec3.1}). Moreover, \citetads{2019ApJ...873....1M} suggested that CMEs on M-dwarfs would be \textit{radio quiet}, due to the strong surface magnetic fields and the correspondingly large Alfv\'en speeds in the corona (see also \citeads{2019ApJ...871..214V}). 

We evaluated these considerations by simulating weakly and partially confined CME events on the archetypical M-dwarf star Proxima Centauri. Results for the former case are presented in Fig.~\ref{Fig_3}. The assumed TD flux-rope eruption parameters were sufficiently large to power the best flare-CME candidate observed in this star so far ($F_{\rm FL}^{\rm X} \simeq 1.7 \times 10^{31}$~erg, $E_{\rm CME}^{\rm K} \simeq~5 \times 10^{31}$~erg; \citeads{2019ApJ...877..105M}). 

Our analysis revealed that due to magnetic suppression, stellar CMEs become super-Alfv\'enic (inducing shocks) at relatively large distances from the stellar surface. Depending on the energy of the escaping CME, our simulations on Proxima Centauri indicate shock locations $5-10$ times higher than the average value registered on the Sun (see Fig.~\ref{Fig_3}). This implies that the plasma density at the shock formation region will be significantly lower, shifting the associated Type~II radio bursts to larger frequencies. The frequency shift is large enough to move the fundamental and first harmonic of the plasma emission to values close to, or even below, the \textit{ionospheric cutoff}\footnote[3]{Usually set at $10$~MHz.} of the Earth. In this way, while CMEs on M-dwarfs are not entirely radio quiet, the surface magnetism of the star and the associated suppression of CMEs create unfavourable conditions for their detection with ground-based instrumentation. 
      
\vspace{-0.5cm}
\section{Concluding Remarks}\label{sec4}
 
Despite its relevance for both high-energy stellar astrophysics and exoplanetary science, eruptive behaviour in stars other than the Sun remains largely unknown. While the vast knowledge of solar CMEs offers a fundamental guideline, its paradigm cannot be followed blindly to the stellar domain. In particular, theoretical and observational considerations have shown that problematic consequences arise from the extension of solar flare-CME relationships to more active stars. In this constraint-limited situation, detailed numerical simulations provide a suitable pathway to study and characterize these energetic phenomena in cool stars. 

As was discussed over the course of this review, our numerical experiments have shown the key role played by the stellar magnetic field on the emerging properties of the CMEs. In particular, state-of-the-art models of CME propagation in active stars have revealed that these transients can be partially and even completely suppressed by the large-scale magnetic field of the star. 

In this way, our investigation indicates that magnetic suppression is a viable mechanism for reducing the flare-CME association rate in active stars. The large-scale field tends to decrease the speed and kinetic energy of the CMEs, a prediction that so far is consistent with the very limited observational data available. Additionally, this process carries with it fundamental consequences for the expected CME signatures in different wavelengths, limiting in some cases our ability to detect these events following solar analogies (i.e. the case of Type II radio~bursts).  

The suppression mechanism can also be extended to a stronger and higher complexity magnetic field regime, such as the one expected for M-dwarfs. While at face value the suppression of CMEs might be taken as beneficial for exoplanets orbiting stars of this spectral type, our simulations show that this actually might not be the case. Both escaping and suppressed CME events could have catastrophic consequences on any exoplanets in the system. These would arise either by the direct CME impact on the exoplanet atmosphere or due to the ionizing radiation from the coronal response induced by the failed eruption. Our models indicate that such CME-driven coronal activity (e.g.~compression flaring, up-flows/down-flows) might be accessible by next-generation high-energy astrophysics instrumentation.

Finally, to account for CMEs (and associated phenomena) in the atmospheric characterization of exoplanets, it is necessary to avoid direct solar extrapolations and instead determine realistic stellar CME properties and occurrence rates. These can only be obtained through a combination of dedicated multi-wavelength observations with appropriate numerical modeling of these energetic events.

\vspace{-0.5cm}
\section*{Acknowledgments}

We would like to thank the referee for constructive feedback. JDAG thanks the organizers of the XMM-Newton Science Workshop 2021 for the invitation to present this investigation. JJD and CG were supported by NASA contract NAS8-03060 to the Chandra X-ray Center and thanks the Director, Pat Slane, for continuing advice and support. OC was supported by NASA NExSS grant NNX15AE05G. KP acknowledges support from the German \textit{Leibniz-Gemeinschaft} under project number P67/2018.












\vspace{-0.5cm}
\bibliography{Biblio}


\end{document}